\documentclass{article} 
\usepackage{iclr2019_conference}
\usepackage[title]{appendix}
\usepackage{array}
\usepackage{times}
\usepackage{latexsym}
\usepackage[hidelinks]{hyperref}
\usepackage{amsmath}
\usepackage{bbm}
\usepackage{amssymb}
\usepackage{algorithm}
\usepackage{algpseudocode}
\usepackage{multirow}
\usepackage{tabu}
\usepackage{booktabs}
\usepackage{graphicx}
\usepackage{subfigure}
\usepackage{booktabs}
\usepackage{breakcites}
\usepackage{enumitem}
\usepackage{tikz}
\usepackage{pgfplots}

\definecolor{g-blue}{HTML}{2E86C1}
\definecolor{g-red}{HTML}{B03A2E}
\definecolor{g-purple}{HTML}{AF7AC5}

\newcolumntype{H}{>{\setbox0=\hbox\bgroup}c<{\egroup}@{}}

\iclrfinalcopy

\title{Navigation-Based Candidate Expansion and Pretrained Language Models for Citation Recommendation}

\author{Rodrigo Nogueira,$^{1,2}$ Zhiying Jiang,$^2$ Kyunghyun Cho,$^{3,4,5,6}$ and Jimmy Lin$^2$\\[0.2cm]
$^1$ Tandon School of Engineering, New York University \\
$^2$ David R. Cheriton School of Computer Science, University of Waterloo \\
$^3$ Courant Institute of Mathematical Sciences, New York University \\
$^4$ Center for Data Science, New York University \\
$^5$ Facebook AI Research~~
$^6$ CIFAR Azrieli Global Scholar \\
}

\begin{document}

\maketitle

\begin{abstract}
Citation recommendation systems for the scientific literature, to help authors find papers that should be cited, have the potential to speed up discoveries and uncover new routes for scientific exploration.
We treat this task as a ranking problem, which we tackle with a two-stage approach:\ candidate generation followed by re-ranking.
Within this framework, we adapt to the scientific domain a proven combination based on ``bag of words'' retrieval followed by re-scoring with a BERT model.
We experimentally show the effects of domain adaptation, both in terms of pretraining on in-domain data and exploiting in-domain vocabulary.
In addition, we introduce a novel navigation-based document expansion strategy to enrich the candidate documents processed by our neural models.
On three different collections from different scientific disciplines, we achieve the best-reported results in the citation recommendation task.
\end{abstract}

\section{Introduction}

The volume of scientific publications is growing at an incredible rate. 
For example, 800,000 papers are added per year to MEDLINE, a database of life sciences and biomedical literature.\footnote{\url{https://www.nlm.nih.gov/bsd/stats/cit_added.html}} 
A recent study estimates that 3M papers are published annually in the English language, with a growth rate of 3--5\% per year~\citep{johnson2018stm}.
This flood of information has made it nearly impossible for researchers to keep abreast of discoveries and innovations, both in their specific sub-field as well as more broadly.
Furthermore, there is an overwhelming amount of material that a scientist entering a new field of study needs to read before becoming familiarized with common concepts, methods, and other foundations.

A number of tools have come along to help researchers cope with this deluge.
For example, keyword-based literature search engines (Google Scholar,\footnote{\url{https://scholar.google.com/}} Microsoft Academic,\footnote{\url{https://academic.microsoft.com/home}} PubMed,\footnote{\url{https://www.ncbi.nlm.nih.gov/pubmed/}} and Semantic Scholar\footnote{\url{https://www.semanticscholar.org/}}) and citation recommendation tools \citep{bollacker1999system,basu2001technical,mcnee2002recommending,kodakateri2009conceptual,he2010context} help scientists find relevant articles, often exploiting citation networks to identify what's important in a particular field.
Methods to automatically populate scientific knowledge bases~\citep{gao2006swan,spangler2014automated,sybrandt2017moliere} form another broad approach to tackling this challenge.

In this work, we investigate the potential of deep language models such as BERT~\citep{devlin2018bert} and large scientific datasets such as Open Research~\citep{ammar2018construction} to improve scientific search tools.
More concretely, we work on the task of scientific literature recommendation, where a paper (title and abstract) is given as a query, and the system's task is to find which papers should be cited.
We use a standard keyword search engine (based on inverted indexes) with BM25 ranking~\citep{robertson1995okapi} to initially retrieve candidate documents and evaluate various pretrained language models as re-rankers.

In domain-specific tasks such as ours, one common issue is low recall in the initial retrieval due to the vocabulary mismatch between query and relevant documents.
As a result, current tools often require from the user multiple rounds of interactions (e.g., query rewrites) where each step requires reading abstracts or passages, a cognitively-demanding task.
We address this challenge by trying to automatically mimic this process to enrich the candidate documents that are fed into the re-ranker, based on the graph of citations.

We find that this simple pipeline comprising off-the-shelf keyword-based initial retrieval, local search in the graph of citations, and BERT re-ranking is more effective than cluster-based methods~\citep{ren2014cluscite,bhagavatula2018content}.
To summarize, our contributions are the following:

\begin{itemize}[leftmargin=*]

\item We introduce a novel method to combine keyword-based retrieval with navigation-based retrieval and obtain state-of-the-art results in three citation recommendation datasets.

\item We evaluate eleven pretrained ranking models and find that pretraining on the target domain and using domain-specific vocabulary lead to large improvements over a general-purpose model.

\item We find that despite the effectiveness of the pretrained language models as query--document relevance estimators, they perform poorly when the term overlap between query and candidate documents is low.
To address this issue, we train with more query--candidate pairs that have low term overlap, but interestingly, such a model performs poorly, even on the training set (see Section~\ref{sec:class_balance}).

\item Contrary to our expectation that query and candidate terms have equal importance to the relevance estimator model, we find that query terms are more important (see Section~\ref{sec:query_type}).

\end{itemize}

\section{Related Work}

Most of the early methods for scientific literature search and recommendation use keyword-based retrieval methods to provide access to the documents~\citep{ginsparg1994first,lawrence1999indexing}.
These methods suffer from the term mismatch problem, which is common in ``bag-of-words'' retrieval methods, but the issue is aggravated by the diversity of the scientific vocabulary~\citep{jerome2001information,dinh2011combining,nabeel2018improved}.
As the number of users grows, popular search engines can exploit interaction signals to learn better ranking models~\citep{mohan2017deep,fiorini2018user,fiorini2018best}.
However, the reported gains are relatively small compared to classic ranking methods such as BM25.

Another common approach in scientific recommendation systems is collaborative filtering~\citep{mcnee2002recommending,liu2015context,chen2018research}.
These methods typically suffer from the cold-start problem, in which there is not enough data about new items (or users) to make predictions accurately.

More recently, cluster-based methods have started to become competitive with traditional retrieval-based methods in this task.
\citet{kanakia2019scalable} cluster papers based on their word embedding representation and use co-citations to alleviate the cold-start problem.
However, they perform human evaluations on a private dataset, which excludes an empirical comparison to our approach.

Navigational methods for information retrieval have been less explored with somewhat limited success.
For example, \citet{nogueira2016end} present an artificial agent trained to navigate Wikipedia to find answers to \textit{Jeopardy!}\ questions.
Similarly, \citet{lao2010relational} use the literature graph and predictive models of proximity measures, such as \textit{Random Walk with Restart}, to recommend papers.
Likewise, navigation on knowledge graphs is extensively used in question-answering tasks~\citep{bordes2014question,das2017go,guu2015traversing,lin2018multi}.

Perhaps closest to our work is \citet{eto2019extended}, who uses a combination of proximity measures from the graph of co-citations to score candidate documents.
The edges in the graph are weighted by the distance in which two citations occur in the citing document. 
This method requires access to the full text of the citing document, which is often not available (for example, due to paywalled content).
Our method, on the other hand, predicts citations using only article abstracts, which are widely available in scientific corpora.

The methods described so far and the one proposed in this work fall in the category of \textit{global} methods, which aim at recommending citations for the entire paper.
Another category comprises \textit{local} methods, which aim at recommending citations for a specific sentence or paragraph in the document~\citep{he2010context,lu2011recommending,huang2012recommending,huang2015neural}.
We do not compare our method with these as we do not assume access to the full text.

\begin{figure*}[t]
\begin{center}
\centerline{\includegraphics[width=0.9\textwidth]{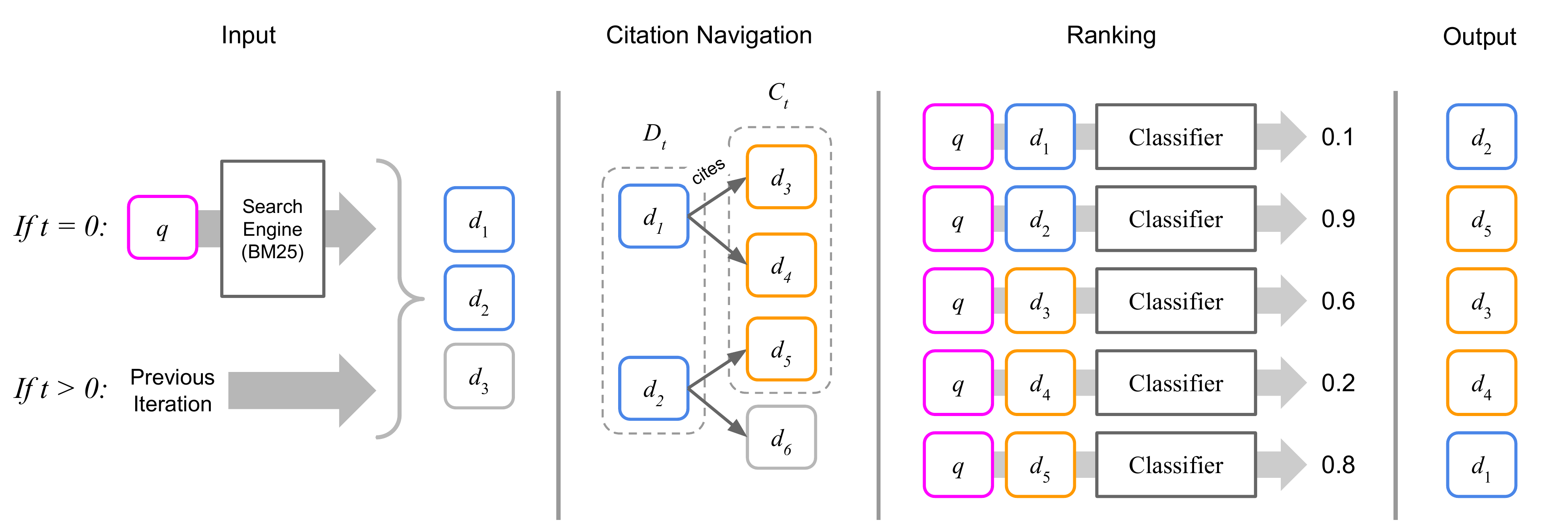}}
\caption{Illustration of our proposed method. The goal of the task is to find papers that should be cited by a given paper $q$, represented as  the concatenation of its title and abstract.
The input at iteration $t=0$ is the top-$k^d_t$ papers $D_t$ retrieved by a keyword search engine when queried with $q$ (in the figure, $k^d_t=2$ and $D_t=(d_1, d_2)$).
For $t>0$, $D_t$ is the output of the previous iteration. In the \textit{navigation} phase, we collect papers $C_t$ cited by $D_t$ (in the figure, $C_t=(d_3, d_4, d_5)$).
We limit the size of $C_t$ to $k^c_t$ by discarding the citations of the lowest-ranked papers in $D_t$ (in the figure, $k^c_t=3$ and $d_6$ is discarded).
In the \textit{ranking} phase, we use a trainable classifier to assign a relevance score with respect to $q$ and each paper in $D_t \cup C_t$ ($d_1$, ..., $d_5$ in the figure).
The output of an iteration of the method comprises the papers re-ranked according to these scores.} \label{fig:overview}
\end{center}
\end{figure*}

\section{Method}

In this work, we address the task of citation recommendation:\ given a partially written paper, the system's task is to return all papers that should be cited in it.
The input query $q$ is the title and abstract of a paper, i.e., we do not have access to the full text. 
We argue that this assumption is crucial to building a useful tool as users might desire recommendations of related papers prior to writing most of the document.

Our method, illustrated in Figure~\ref{fig:overview}, is a multi-stage ranking pipeline that comprises two main phases, \textit{Navigation} and \textit{Ranking}, which can be repeated for multiple iterations $t<T, T \in \mathbb{N}$, where $T$ is a hyperparameter.

In the first iteration ($t=0$), the input to the method is the top-$k^d_t$ papers $D_t$ retrieved by a keyword search engine when queried with query $q$.
For $t>0$, the input is the top-$k^d_t$ papers $D_{t}$ returned by the previous iteration.

\smallskip \noindent {\bf Navigation:}  Given input $D_t$, we collect papers $C_t$ cited by $D_t$. We limit the number of papers in $C_t$ to $k^c_t$ by incrementally removing citations from the lowest-ranked papers in $D_t$. Both $ k^d_t, k^c_t \in \mathbb{N}^{+}$ are hyperparameters of our method. The output of this phase is $D_t \cup C_t$.

\smallskip \noindent {\bf Ranking:} We compute the probability $p(d^i_t|q)$ for each paper $d^i_t \in D_t \cup C_t$ being relevant to $q$.
For this, we use a BERT~\citep{devlin2018bert} re-ranker model from~\citet{nogueira2019passage}.
Using the same notation as Devlin et al., we feed the query tokens as sequence A and the candidate paper tokens as sequence B.
In our task, both the query and the candidate paper are the concatenation of the title and abstract of each paper, resulting in an input that is often longer than the maximum tokens allowed by the model, which is typically 512 tokens. To circumvent this, we iteratively remove tokens from the largest sequence until the maximum number of tokens is reached. We use the model as a binary classifier:\ we feed the [CLS] vector to a single layer neural
network to obtain $p(d^i_t|q)$. The output of an iteration of our method is a list of papers $D_t \cup C_t$ ranked by $p(d^i_t|q)$. Training details are provided in Section~\ref{sec:reranker_training}.

Intuitively, our method tries to mimic a common user behavior:\ an initial set of candidate papers is refined by repeatedly gathering references from it and selecting the ones that deserve further reading.
When multiple iterations are applied, the \textit{navigation} phase allows exploration of distant literature while the \textit{ranking} phase avoids an exponential growth of retrieved papers by keeping only the ones estimated as most relevant.

From another perspective, our method is similar to the beam search navigational algorithm proposed by \citet{nogueira2016end}. Ours, however, has broader applicability as we do not need a hierarchy of links (such as the ones in Wikipedia) to start navigating.
Instead, we use keyword search to retrieve an initial set of candidate documents.

\section{Experimental Setup}

\begin{table}[t]
\begin{center}
\begin{tabular}{lrrr}
 & Open Research & DBLP & PubMed \\
\noalign{\vskip 1mm}
\toprule
\noalign{\vskip 1mm}
Total \# of docs & 6,892,252 & 50,227 & 47,347\\
Total \# of citations & 44,400,729 & 156,807 & 825,371\\
Avg.\ \# citations per doc & 6.45 & 3.12 & 17.43\\
Avg.\ len.\ per doc (char) & 1,391 & 1,193 & 1,504\\
\noalign{\vskip 0.5mm}
\midrule
\noalign{\vskip 0.5mm}
Queries - Train & 3,343,809 & 27,322 & 26,793\\
\kern 11.8mm - Dev & 487,582 & 8,324 & 2,768\\
\kern 11.8mm - Test & 464,449 & 931 & 8,815\\
\noalign{\vskip 0.5mm}
\midrule
\noalign{\vskip 0.5mm}
q/rel.\ doc pairs - Train & 32,470,673 & 106,011 & 558,674\\
\kern 21.8mm - Dev & 5,985,787 & 38,628 & 66,655\\
\kern 21.8mm - Test & 5,944,269 & 12,168 & 200,042\\
\noalign{\vskip 1mm}
\midrule
\end{tabular}
\end{center}
\vspace{-2mm}
\caption{Statistics of the datasets.}
\label{tab:dataset_summary}
\end{table}

\subsection{Datasets}

\smallskip \noindent {\bf Open Research.}
We train and evaluate our models on the Open Research corpus~\citep{ammar2018construction}.\footnote{\url{https://s3-us-west-2.amazonaws.com/ai2-s2-research-public/open-corpus-archive/2017-02-21/papers-2017-02-21.zip}}
It comprises 7.2M computer science and biomedical paper abstracts and their references.
We closely follow the data processing from \citet{bhagavatula2018content} to create the training, development, and test sets. That is, we sort papers by publication year and use the oldest 80\% for training (1991--2014), the next 10\% for development (2014--2015), and the most recent 10\% for testing (2015--2016).
Since the development and test sets are too large (400k+ papers), we randomly sample 20k examples from each set.
We remove papers that do not cite any others or that have no year of publication.
Finally, we remove citations of papers that are not in the corpus or whose publication year is less than that of the citing paper.
Table~\ref{tab:dataset_summary} shows the statistics of the resulting dataset.

Note that our dataset statistics do not match the ones reported in \citet{bhagavatula2018content}, but they match those output by the evaluation script provided by the authors.\footnote{\url{https://github.com/allenai/citeomatic/blob/master/citeomatic/scripts/evaluate.py}} The difference is that the authors report in the paper statistics before the filtering steps (such as removing papers without references).
Thus, our corpus and dataset splits match exactly the ones used by the authors to run their experiments.

\smallskip \noindent {\bf DBLP and PubMed.}
The DBLP and PubMed datasets are introduced by \citet{ren2014cluscite} and comprise papers from computer science and biomedicine, respectively. We apply the same data processing steps from~\citet{bhagavatula2018content}, and the resulting datasets are summarized in Table~\ref{tab:dataset_summary}. 

\subsection{Deduplication}

When evaluating our method on DBLP and PubMed, we use models trained on Open Research's training set as this yields better results than training on the much smaller DBLP and PubMed training sets.
To avoid leaking training data into the evaluation sets, we use the following method to remove documents in Open Research's training set that appear in the development and test sets of PubMed and DBLP. We remove special characters from the title and use Jaccard similarity to calculate the closeness of two documents.
We set our shingles to a single word and our threshold to 0.7. This method results in approximately half of the papers in the development and test sets of PubMed and DBLP being removed from the training set of Open Research.

\subsection{Re-ranker Training}
\label{sec:reranker_training}

To obtain the positive and negative examples used to train our binary classification models, we retrieve the top 10 papers for a query using the Anserini IR toolkit\footnote{\url{http://anserini.io/}}~\citep{Yang_etal_SIGIR2017,Yang_etal_JDIQ2018} with BM25 ranking.
Among these, less than 10\% on average are relevant papers (positive examples).
We do not balance positive and negative examples, and we discuss this decision in Section~\ref{sec:multiple_iterations}.

We start training from a pretrained BERT model and fine-tune it to our task using cross-entropy loss:
\begin{equation}
L = - \sum_{j \in J_{\text{pos}}} \log(p(d_j|q)) - \sum_{j \in J_{\text{neg}}} \log (1 - p(d_j|q)),
\end{equation}
where $J_{\text{pos}}$ and $J_{\text{neg}}$ are the sets of indexes of the relevant and irrelevant papers, respectively, and $p(d_j|q)$ is the relevance probability the model assigned to $j$-th paper.

We fine-tune the model using Google's TPUs
v3 with a batch size of 128 (128 sequences $\times$ 512 tokens = 65,536 tokens/batch)\ for 300k iterations, which takes approximately three days. This corresponds to training on 38.4M (300k $\times$ 128) query--candidate pairs, or 1.1 epochs. We do not see any improvements in the development set when training for another 700k iterations, which is equivalent to 3.8 epochs.

We use Adam~\citep{kingma2014adam} with the initial learning rate set to $3 \times 10^{-6}$, $\beta_1 = 0.9$, $\beta_2 = 0.999$, L2 weight decay of 0.01, learning rate warmup over the first 10,000 steps, and linear decay of the learning rate. We use a dropout probability of $0.1$ in all layers.

\subsection{Inference and Metrics}

At inference time, we first retrieve the top 1000 candidate documents with the title and abstract as the query using BM25 ranking in Anserini.
These documents are optionally further expanded with the proposed navigation method and re-ranked with SciBERT-Large.
Following \citet{bhagavatula2018content}, we evaluate the models using $\text{F}_1$ of the top 20 retrieved papers ($\text{F}_1$@20) and Mean Reciprocal Ranking (MRR) of top 1000 retrieved papers.
We additionally report Recall@1000 to compare the effectiveness of our navigation phase with keyword-based search.

\section{Results}

Our main results are shown in Tables~\ref{tab:result_openresearch}, \ref{tab:result_dblp}, and \ref{tab:result_pubmed}.
Our ranking model, SciBERT-Large, is selected based on the experiments in Section~\ref{sec:pretraining}.
On the Open Research dataset, our best configuration (BM25 + Navigation + ranking SciBERT-Large) improves upon the best previous result by more than 2 points on $\text{F}_1$@20 and 11 points on MRR.
On the smaller DBLP and PubMed datasets, our best method is on par with the state of the art. 
Note that our BERT-based models are trained only on Open Research as we achieve better results than training on the smaller datasets.

\begin{table*}[t]
\centering\centering\resizebox{1.0\textwidth}{!}{
\begin{tabular}{l|cc|cc|cc}
 & \multicolumn{2}{c}{$\text{F}_1$@20} & \multicolumn{2}{c}{MRR} & \multicolumn{2}{c}{R@1000}\\
 & Dev & Test& Dev & Test & Dev & Test\\
\noalign{\vskip 1mm}
\toprule
\noalign{\vskip 1mm}
BM25~\citep{bhagavatula2018content} & - & 0.058 & - & 0.218 & - & -\\
BM25 (Anserini, Ours) & 0.082 & 0.089 & 0.279 & 0.312 & 0.424 & 0.421\\
\noalign{\vskip 1mm}
\midrule
\noalign{\vskip 1mm}
Citeomatic~\citep{bhagavatula2018content} & - & 0.125 & - & 0.330 & - & -\\
BM25 + Ranking SciBERT-Large (Ours) & 0.136 & 0.132 & 0.430 & 0.431 & 0.424 & 0.421\\
BM25 + Navigation + Ranking SciBERT-Large (Ours) & \textbf{0.154} & \textbf{0.148} & \textbf{0.451} & \textbf{0.445} & \textbf{0.658} & \textbf{0.624}\\
\end{tabular}
}
\vspace{-2mm}
\caption{Main results on Open Research.}
\label{tab:result_openresearch}
\vspace{4mm}
\end{table*}

\begin{table*}[t]
\centering\centering\resizebox{1.0\textwidth}{!}{
\begin{tabular}{l|cc|cc|cc}
 & \multicolumn{2}{c}{$\text{F}_1$@20} & \multicolumn{2}{c}{MRR} & \multicolumn{2}{c}{R@1000}\\
 & Dev & Test& Dev & Test & Dev & Test\\
\noalign{\vskip 1mm}
\toprule
\noalign{\vskip 1mm}
BM25~\citep{bhagavatula2018content} & - & 0.119 & - & 0.425 & - & -\\
BM25 (Anserini, Ours) & 0.105 & 0.194 & 0.352 & 0.585 & 0.669 & 0.691 \\
\noalign{\vskip 0.5mm}
\midrule
\noalign{\vskip 0.5mm}
ClusCite~\citep{ren2014cluscite} & - &  0.237 & - & 0.548 & - & -\\
Citeomatic~\citep{bhagavatula2018content} & - & \textbf{0.303} & - & 0.689 & - & -\\
\noalign{\vskip 0.5mm}
\midrule
\noalign{\vskip 0.5mm}
BM25 + Ranking SciBERT-Large (Ours) & 0.149 & 0.272 & 0.472 & \textbf{0.714} & 0.669 & 0.691 \\
BM25 + Navigation + Ranking SciBERT-Large (Ours) & 0.148 & 0.277 & 0.469 & \textbf{0.714} & \textbf{0.817} & \textbf{0.862} \\
\end{tabular}
}
\vspace{-2mm}
\caption{Main results on DBLP.}
\label{tab:result_dblp}
\vspace{4mm}
\end{table*}

\begin{table*}[t]
\centering\centering\resizebox{1.0\textwidth}{!}{
\begin{tabular}{l|cc|cc|cc}
 & \multicolumn{2}{c}{$\text{F}_1$@20} & \multicolumn{2}{c}{MRR} & \multicolumn{2}{c}{R@1000}\\
 & Dev & Test& Dev & Test & Dev & Test\\
\noalign{\vskip 1mm}
\toprule
\noalign{\vskip 1mm}
BM25~\citep{bhagavatula2018content} & - & 0.209 & - &0.574 & - & -\\
BM25 (Anserini, Ours) & 0.299 & 0.268 & 0.793 & 0.721 & 0.794 & 0.765 \\
\noalign{\vskip 0.5mm}
\midrule
\noalign{\vskip 0.5mm}
ClusCite~\citep{ren2014cluscite} & - & 0.274 & - & 0.578 & - & -\\
Citeomatic~\citep{bhagavatula2018content} & - & \textbf{0.329} & - & 0.771 & - & -\\
\noalign{\vskip 0.5mm}
\midrule
\noalign{\vskip 0.5mm}
BM25 + Ranking SciBERT-Large (Ours) & 0.326 & 0.304 & 0.835 & \textbf{0.792} & 0.794 & 0.765 \\
BM25 + Navigation + Ranking SciBERT-Large (Ours) & 0.324 & 0.301 & 0.836 & \textbf{0.790} & \textbf{0.903} & \textbf{0.876}\\
\end{tabular}
}
\vspace{-2mm}
\caption{Main results on PubMed.}
\label{tab:result_pubmed}
\end{table*}

Our baseline BM25 implementation is 3--7 points higher in $\text{F}_1$@20 than the implementation of Bhagavatula et al.
This is due to the choice of query form, which is analyzed in Section~\ref{sec:query_type}, and perhaps a better implementation of BM25 in Anserini.
Our method without the \textit{navigation} component (BM25 + Ranking SciBERT-Large) is at least on par with the state-of-the-art method in this task (Citeomatic).
By including documents from navigation (BM25 + Navigation + Ranking SciBERT-Large), we increase Recall@1000 by 10--20 points and $\text{F}_1$@20 by 1--2 points.
The smaller improvement in $\text{F}_1$@20 compared to recall is unexpected, and we investigate this in Section~\ref{sec:multiple_iterations}.

Our method appears to be as effective and more scalable than a cluster-based approach.
For example, \citet{bhagavatula2018content} requires at least 100 GB of RAM to search the 7M documents in the Open Research corpus,\footnote{\url{https://github.com/allenai/citeomatic\#citeomatic-evaluation}} whereas keyword search has far more modest memory requirements.

In the next sections, we investigate the effectiveness of our method by evaluating various pretrained language models, as well as the effects of multiple iterations, class imbalance, and different queries.

\begin{table*}[t]
\centering\centering\resizebox{1.0\textwidth}{!}{
\begin{tabular}{cl|ccccc|cc}
& Pretrained Model & Size & Pretraining Corpus & Tokens & Vocabulary & Cased & $\text{F}_1$@20 & MRR\\
\noalign{\vskip 1mm}
\midrule
\noalign{\vskip 1mm}

(1) & NCBI & Base & PubMed+MIMIC & 4.5B & Wiki+Books &  & 0.093 & 0.315 \\
(2) & NCBI & Large & PubMed+MIMIC & 4.5B & Wiki+Books &  & 0.105 & 0.352 \\
\noalign{\vskip 1mm}
\midrule
\noalign{\vskip 1mm}
(3) & Google & Base & Wiki+Books & 3.3B & Wiki+Books &  & 0.113 & 0.374 \\
(4) & Google & Large & Wiki+Books & 3.3B & Wiki+Books &  & 0.115 & 0.373 \\
(5) & Google WWM & Large & Wiki+Books & 3.3B & Wiki+Books &  & 0.121 & 0.399 \\
\noalign{\vskip 1mm}
\midrule
\noalign{\vskip 1mm}
(6) & RoBERTa & Large & Various (Non-Scientific) & 33B & (Non-Scientific) & & 0.125 & 0.409 \\
\noalign{\vskip 1mm}
\midrule
\noalign{\vskip 1mm}
(7) & BioBERT v1.1 & Base & Wiki+Books+PubMed+PMC & 21.3B & PubMed+PMC & \checkmark & 0.128 & 0.417 \\
\noalign{\vskip 1mm}
\midrule
\noalign{\vskip 1mm}
(8) & SciBERT & Base & Open Research (1M Full Papers) & 3.2B & Wiki+Books &  & 0.125 & 0.409 \\
(9) & SciBERT & Base & Open Research (1M Full Papers) & 3.2B & Open Research &  & 0.131 & 0.423 \\
(10) & SciBERT (ours) & Large & Open Research (7M Abstracts) & 1.4B &  Wiki+Book &  & 0.135 & 0.420 \\
(11) & SciBERT (ours) & Large & Open Research (7M Abstracts) & 1.4B & Open Research &  & \textbf{0.137} & \textbf{0.430} \\
\end{tabular}
}
\vspace{-2mm}
\caption{Results on Open Research's development set of BERT-based models pretrained under different settings. All models are fine-tuned for approximately one epoch on the training set.}
\label{tab:pretraiing}
\end{table*}

\subsection{In- vs. out-domain pretraining corpus}
\label{sec:pretraining}

Here we investigate how different pretraining configurations change effectiveness in the target task.
The results, shown in Table~\ref{tab:pretraiing}, are from fine-tuning the pretrained models on Open Research's training set for 300k iterations with a batch of size 128, which corresponds to approximately 1.1 epochs.
In the remainder of this paper, we call \textit{in-domain} corpus a collection whose majority of documents are of the same domains as those in Open Research (i.e., biomedicine and computer science), and we call \textit{out-domain} corpus a collection whose majority of papers are not from those domains.

The models pretrained on an in-domain corpus, i.e., BioBERT ~\citep{lee2019biobert} (row 7) and SciBERT~\citep{beltagy2019scibert} (rows 8--11), give significant improvements in the target task over models pretrained on a corpus of a similar size but different domain (rows 3--5).
Pretraining on an out-domain corpus ten times the size of the in-domain corpus results in lower effectiveness on the target task, RoBERTa~\citep{liu2019roberta} (row 6 vs.\ 10).
We conclude that, at least for the task of citation recommendation, pretraining on a smaller in-domain corpus is more effective than pretraining on a larger but out-domain corpus.

When pretraining settings are kept the same except for the vocabulary, the use of in-domain vocabulary gives 5--10\% improvement over out-domain vocabulary (row 8 vs.\ 9 and row 10 vs.\ 11).
\citet{beltagy2019scibert} report a similar finding in other tasks. 

NCBI models~\citep{peng2019transfer} (rows 1 and 2) are pretrained on an in-domain corpus but produce worse results than models pretrained on an out-domain corpus of a similar size (rows 3--5).
They also underperform when compared to SciBERT-Base (row 8), which is pretrained on an in-domain corpus of a similar size but comprises full papers instead of abstracts.
As noted by \citet{beltagy2019scibert}, this result indicates that pretraining with longer documents improves the target task effectiveness. 

We find that model size is even more important than document length; our SciBERT-Large models (rows 10 and 11) have higher effectiveness than the SciBERT-Base models (rows 8 and 9) despite being pretrained on a smaller corpus of 7M paper abstracts (1.4B tokens) as opposed to 1M full-text papers (3.2B tokens).

\subsection{Multiple iterations}
\label{sec:multiple_iterations}

We evaluate our method with up to three iterations ($T=3$) of \textit{navigation} and \textit{ranking}.
The hyperparameters $k^d_t$ and $k^c_t$ are found by setting $k^d_t + k^c_t = 1000$, sweeping $k^d_t$ over $\{0, 100, 200, ..., 1000\}$, and using the values that yield the highest R@1000 on the development set of Open Research. These values are: $k^d_0 = 300$ ($k^c_0 = 700$), $k^d_1 = 700$ ($k^c_1 = 300$), and $k^d_2 = 900$ ($k^c_2 = 100$).

\begin{table}[t]
\begin{center}
\begin{tabular}{l|cccc}  
Iterations& $\text{F}_1$@20 & MRR & R@1000 & Term Overlap\\
\noalign{\vskip 1mm}
\midrule
\noalign{\vskip 1mm}
0 Iteration (no navigation) & 0.131 & 0.423 & 0.423 & 0.146 \\
1 Iteration & \textbf{0.151} & \textbf{0.449} & 0.658 & 0.126\\
2 Iterations & 0.148 & 0.446 & \textbf{0.681} & 0.120\\
3 Iterations & 0.147 & 0.445 & 0.681 & 0.118\\
\end{tabular}
\end{center}
\vspace{-2mm}
\caption{Results on Open Research's development set when varying the number of iterations $T$, with SciBERT-Base as the re-ranker.
The first row uses the output of BM25 as the input to the re-ranker (i.e., no citation navigation).
The last column (Term Overlap) shows the overlap between query and candidate document terms, excluding stopwords.}
\label{tab:nav_steps}
\vspace{4mm}
\end{table}

We show the development set results in Table~\ref{tab:nav_steps}.
Although Recall@1000 increases with more iterations, $\text{F}_1$@20 and MRR reach their peak with one iteration.
We conjecture that this is due to a limitation of our current re-ranker, whose effectiveness drops when candidate papers that have less lexical similarity (i.e., term overlap) with the query are presented at inference time. 
To support this claim, we empirically verify that the proportion of candidate document terms (excluding stopwords) that overlaps with the query (last column of Table~\ref{tab:nav_steps}) decreases by quite a bit as we go from one to two iterations, hence the drop in $\text{F}_1$@20 and MRR despite the increase in recall.

\subsection{Class Imbalance}
\label{sec:class_balance}

Because we only use the top 10 papers returned by BM25 as training examples, the BERT-based models in this work are trained with more negative examples than positive ones (94\% vs.~6\%).
In a separate experiment, to balance these classes, we include in the training phase pairs of query and relevant papers not retrieved by BM25, but this results in $\text{F}_1$@20 and MRR close to zero in both training and development sets. We obtain a similar result when adding to the training set negative candidates randomly sampled from the corpus.
We hypothesize that, although BERT is a strong model for the document ranking task, it partly relies on exact term match to learn relevance. 
Thus, when we sample training documents {\it not} using an exact term match method such as BM25, fewer terms between the query and the candidate paper match, making learning harder. Further studies should investigate if this limitation applies to other tasks as well.

\begin{table*}
\centering\centering\resizebox{1.0\textwidth}{!}{
\begin{tabular}{l|ccc|ccc|ccc}
& \multicolumn{3}{c|}{Open Research} & \multicolumn{3}{c|}{PubMed} & \multicolumn{3}{c}{DBLP} \\
Query Type & $\text{F}_1$@20 & MRR & R1000 & $\text{F}_1$@20 & MRR & R1000 & $\text{F}_1$@20 & MRR & R1000\\
\noalign{\vskip 1mm}
\midrule
\noalign{\vskip 1mm}
Key Terms (Whoosh) & 0.065 & 0.251 & 0.282 & 0.201 & 0.595 & 0.604 & 0.130 & 0.425 & 0.510 \\
Title & 0.063 & 0.244 & 0.287 & 0.199 & 0.584 & 0.654 & 0.133 & 0.424 & 0.551\\
Title and Abstract & \textbf{0.095} & \textbf{0.351} & \textbf{0.363} & \textbf{0.268} & \textbf{0.720} & \textbf{0.765} & \textbf{0.194} & \textbf{0.585} & \textbf{0.691}\\
\end{tabular}}
\caption{BM25 results on Open Research's development set when different query types are used. Navigation and BERT-based re-ranking are not applied.}
\label{tab:query}
\end{table*}

\subsection{Query Analysis}
\label{sec:query_type}

In the citation recommendation task, the query can take many forms, such as the title of the paper, the concatenation of title and abstract, or keywords extracted from the text.
Here we investigate how these query types affect the effectiveness of a keyword-based retrieval method.

In Table~\ref{tab:query}, we show the effectiveness of BM25 on the Open Research development set.
For \textit{Key Terms}, we follow \citet{bhagavatula2018content} and use Whoosh\footnote{\url{https://whoosh.readthedocs.io/en/latest/}} to first create an index and then extract key terms from the title and abstract with Whoosh's \texttt{key\_terms\_from\_text} method.
Despite being faster due to having fewer query terms, the results show that this method has lower effectiveness than simply concatenating the title and abstract of the paper.

One of the limitations of transformer-based models (including BERT) is that memory consumption increases quadratically with the number of tokens in the input sequence.
On modern hardware such as TPUs v3 or GPU V100s, the maximum number of tokens that one can efficiently train a BERT-Large model is approximately 512.
In our task, since the concatenation of query and candidate tokens is typically longer than this value, there is a trade-off between the number of tokens we allocate for each sequence type.

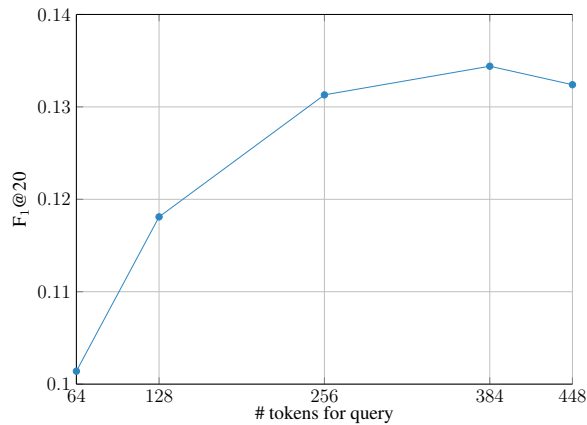
\begin{figure}
\centering
\begin{tikzpicture}[scale = 0.6]
\begin{axis}[
width=0.9\columnwidth,
height=0.70\columnwidth,
legend cell align=left,
legend style={font=\tiny},
mark options={mark size=3},
font=\large,
axis y line*=left,
xmin=64, xmax=448,
ymin=0.10, ymax=0.14,
log ticks with fixed point,
xtick={64, 128, 256, 384, 448},
ytick={0.1, 0.11, 0.12, 0.13, 0.14},
legend pos=south east,
xmajorgrids=true,
ymajorgrids=true,
xlabel style={font = \large, yshift=1ex},
xlabel=\# tokens for query,
ylabel=$\text{F}_1$@20,
ylabel style={font = \large, yshift=0ex}]
    \addplot[mark=*,g-blue, mark options={scale=1}] plot coordinates {
    (64, 0.1014)(128, 0.1181)(256, 0.1313)(384, 0.1344)(448, 0.1324)
    };
\end{axis}
\end{tikzpicture}
\caption{$\text{F}_1$@20 on the development set when varying the number of tokens allocated to the input sequence (whose limit is 512 tokens) for the query (as opposed to the candidate document).}
\label{fig:query_candidate_tokens}
\end{figure}

In Figure~\ref{fig:query_candidate_tokens}, we show how effectiveness changes as we allocate more tokens to the query than to the candidate document while limiting the sum of the two sequence types to 512 tokens.
These results are obtained with BM25 + SciBERT-Base.
The curve shows that query terms are more important to the re-ranker model, as increasing query tokens from 64 to 256 increases $\text{F}_1$@20 by 2 points.
Decreasing candidate document tokens from 256 to 64 barely changes $\text{F}_1$@20.
This result is somewhat surprising as one expects the two sequences to have equal importance in the task of query--document relevance estimation.
Note that in all previous experiments (Tables~\ref{tab:result_openresearch}--\ref{tab:nav_steps}), we used 256 tokens for the query and 256 for the candidate; this suggests that our main results might be even higher had we tuned this hyperparameter as well.
Future work should investigate if this is particular to citation recommendation, or if it also occurs in other retrieval tasks with lengthy queries.

\section{Conclusion}

We provide an extensive evaluation of pretrained language models for the scientific literature recommendation task.
We find that in-domain pretraining and domain-specific vocabulary greatly improve effectiveness.
Local search in the graph of citations significantly mitigates the vocabulary mismatch problem due to ``bag of word'' initial retrieval:\ recall increases by 10--20 points in three datasets.

Additionally, we present two unexpected findings:

\begin{enumerate}[leftmargin=*]

\item The effectiveness of BERT-based models degrades as the term overlap between query and candidate document decreases; increasing the number of training examples that have low query--candidate term overlap results in a poor model (i.e., does not fix this issue).
This finding suggests that, despite the wealth of semantic knowledge captured by BERT-based models, they still rely to a large degree on exact term matches for this task.

\item Despite the symmetry of the two inputs when trying to estimate the relevance of a candidate article to a query article, we find that terms from the query article are more important than terms from the candidate article in allocating ``space'' for BERT input.

\end{enumerate}

Future work should investigate these observations.

\section*{Acknowledgments}

We would like to thank Google for Google Cloud credits.

\bibliographystyle{iclr2019_conference}
\bibliography{main}

\end{document}